\documentclass[sigconf]{acmart}
\AtBeginDocument{%
  \providecommand\BibTeX{{%
    \normalfont B\kern-0.5em{\scshape i\kern-0.25em b}\kern-0.8em\TeX}}}

\usepackage{makecell}
\usepackage{longtable}
\usepackage{caption}
\newcolumntype{L}[1]{>{\raggedright\let\newline\\\arraybackslash\hspace{0pt}}m{#1}}
\newcolumntype{C}[1]{>{\centering\let\newline\\\arraybackslash\hspace{0pt}}m{#1}}
\newcolumntype{R}[1]{>{\raggedleft\let\newline\\\arraybackslash\hspace{0pt}}m{#1}}

\usepackage{enumitem}
\usepackage{xcolor}
\usepackage{multirow}
\usepackage{siunitx}

\newcommand{\framework}[0]{\textit{Parameters of Participation}}

\copyrightyear{2023} 
\acmYear{2023} 
\setcopyright{rightsretained} 
\acmConference[EAAMO '23]{Equity and Access in Algorithms, Mechanisms, and Optimization}{October 30-November 1, 2023}{Boston, MA, USA}
\acmBooktitle{Equity and Access in Algorithms, Mechanisms, and Optimization (EAAMO '23), October 30-November 1, 2023, Boston, MA, USA}
\acmDOI{10.1145/3617694.3623261}
\acmISBN{979-8-4007-0381-2/23/11}

\acmConference[EAAMO '23]{EAAMO '23: ACM conference on Equity and Access in Algorithms, Mechanisms, and Optimization}{October 30--November 01, 2023}{Boston, MA}
\acmBooktitle{EAAMO '23:ACM conference on Equity and Access in Algorithms, Mechanisms, and Optimization, Boston, MA}
\acmISBN{979-8-4007-0381-2/23/11}

\begin{document}

\title[The Participatory Turn in AI Design]{The Participatory Turn in AI Design: Theoretical Foundations and the Current State of Practice}

\author{Fernando Delgado}
\authornote{The first two and last two authors contributed equally to this research.}
\affiliation{%
  \institution{Cornell University}
  \city{Ithaca}
  \state{New York}
  \country{USA}}
  \email{fad33@cornell.edu}

\author{Stephen Yang}
\authornotemark[1]
\authornote{Stephen began this work while at Cornell University, and Michael began this work while at Microsoft Research.}
\affiliation{%
  \institution{University of Southern California}
  \city{Los Angeles}
  \state{California}
  \country{USA}}
\email{stepheny@usc.edu}

\author{Michael Madaio}
\authornotemark[1]
\authornotemark[2]
\affiliation{%
\institution{Google Research}
\city{New York}
\state{New York}
\country{USA}
\email{madaiom@google.com}
}

\author{Qian Yang}
\authornotemark[1]
\affiliation{%
  \institution{Cornell University}
  \city{Ithaca}
  \state{New York}
  \country{USA}
}
\email{qianyangcornell.edu}

\renewcommand{\shortauthors}{Delgado et al.}

\begin{abstract}
  
Despite the growing consensus that stakeholders affected by AI systems should participate in their design, enormous variation and implicit disagreements exist among current approaches. 
For researchers and practitioners who are interested in taking a participatory approach to AI design and development, it remains challenging to assess the extent to which any participatory approach grants substantive agency to stakeholders. 
This article thus aims to ground what we dub the ``\textit{participatory turn}'' in AI design by synthesizing existing theoretical literature on participation and through empirical investigation and critique of its current practices. 
Specifically, we derive a conceptual framework through synthesis of literature across technology design, political theory, and the social sciences that researchers and practitioners can leverage to evaluate approaches to participation in AI design.
Additionally, we articulate empirical findings concerning the current state of participatory practice in AI design based on an analysis of recently published research and semi-structured interviews with 12 AI researchers and practitioners.
We use these empirical findings to understand the current state of participatory practice and subsequently provide guidance to better align participatory goals and methods in a way that accounts for practical constraints.


\end{abstract}

\ccsdesc[500]{Human-centered computing~Participatory design}
\ccsdesc[500]{Human-centered computing~Interaction design theory, concepts and paradigms}
\ccsdesc[500]{Human-centered computing~Computer supported cooperative work}



\keywords{Artificial intelligence, machine learning, participation, participatory design, power} 


\maketitle



\section{Introduction}

As artificial intelligence (AI) systems are developed and deployed across various sectors (e.g., hiring, healthcare, education, content moderation), there have been increasing calls to involve members of communities impacted by AI systems in their design \cite{icml2020wkshp,informs2021wkshp,loi2018pd,cscw2018wkshp,pair2020boundary, li_2021,pair2021chi, hoffmann2020terms, loi2019co, amershi2019guidelines, OpenAI_2023}. In part, such calls for participation in AI design argue that participation can enable AI systems to better reflect the values, preferences, and needs of users and other impacted stakeholders, or more broadly, that participation will empower stakeholders in shaping the design of AI systems \cite{Birhane2022power, algo_2021,deloitte2018participatory, loi2019co, denton2020bringing, shneiderman2021human,amershi2019guidelines, baumer2017toward,bondi2021envisioning, shen2021value,cheng2019explaining,shen2020designing, martin2020participatory, krafft2021facct, schiff2020principles}.
\looseness=-1 

However, despite a growing consensus that stakeholders \textit{should} participate more in AI design, there is enormous variation in the methods and theories applied to achieve that participation, even with respect to the goals for leveraging participation in the first place. 
There are, for example, practitioners that take a community-centric approach in defining and operationalizing stakeholder participation upstream of any system or product design activity \cite{katell2020facct}; community collectives of African NLP researchers \cite{nekoto2020participatory} and queer AI researchers \cite{queerinai2023queer} have formed to design and evaluate algorithmic systems that reflect their communities' needs.
In a commercial context, industry practitioners argue that participation helps develop products that better align with people's different wants and needs and that it accelerates innovation \cite{seger2023democratising} while improving profitability \cite{groves2023going}.
Further downstream from a traditional design phase, many practitioners involve stakeholders as the ``human infrastructure''~\cite{mateescu2019ai} or ``data labor''~\cite{miceli2022data} that enables ML systems to work.
This involves including stakeholders in data generation and interactions with data collection infrastructures \cite{mateescu2019ai,cukier2014rise, barnett2022crowdsourcing}, data labeling or annotation \cite{hovy2013learning,miceli2020between, patton2020contextual, suresh2022towards}, feedback for optimization~\cite{casper2023open, liu2023perspectives, bai2022training, ganguli2022red}, and end users’ discretionary work to make the system work \cite{alkhatib2019street}.
Meanwhile, some researchers have begun to leverage language models to generate synthetic user research data positing such data as suitable ``proxies for understanding human experiences'' \cite{argyle2023, hamalainen2023evaluating}.

As large-scale AI models such as large language models are trained on ever larger swaths of human-generated data \cite[e.g.][]{dodge2021documenting} and ``fine-tuned'' through processes involving human feedback in some form \cite[e.g.,][]{bai2022training}, the stakes are high for what it means for stakeholders to meaningfully participate in shaping the design of AI---and what it might mean for an AI system or its design process to be branded as ``participatory AI,'' with all of the implications for empowering stakeholders that that term implicitly conveys. While diversity in approach is warranted to address the various opportunities and constraints present in any system design context, the current state of participatory AI speaks to a largely unaccounted for and unaccountable heterogeneity that lacks a shared ethos or set of principles. 

To address this gap, we attempt in this paper to provide an analytic structure and empirical foundation to help mature participatory practice in AI design. To do that, we first synthesize theories and approaches to participation across multiple fields (e.g., technology design, policy and governance, international development) to develop a conceptual framework that enables us to articulate key distinctions across a broad spectrum of participatory strategies and tactics.  
We then use this conceptual framework to analyze a corpus of 80 research papers that describe their approach as ``participatory AI.'' In mapping this corpus to our conceptual framework, we map out the contemporary terrain of practice. Further, we interviewed 12 authors of papers in this corpus to understand the motivations, challenges, and aspirations of researchers leveraging participatory AI. 

We find that most current participatory AI efforts \textit{consult} stakeholders for input on discrete implementation parameters, rather than \textit{empower} them to make key AI design decisions. In addition, we find that researchers and practitioners implementing participatory AI projects experience a tension between their aspirations to empower participants and the practical constraints (e.g., resources, timelines) they faced in attempting to do so. To resolve these tensions, we find that participatory AI projects adopt tactics such as relying on \textit{proxies for participation}---including both human and algorithmic proxies---to represent stakeholders in shaping AI systems. Such tactics raise important questions regarding the affordances and limits of simulated stand-ins in the context of AI design. 

This paper makes three contributions. First, we provide a conceptual framework for analyzing participatory approaches to AI design in a way that connects goals to techniques. Second, we contribute empirical findings about motivations and challenges within the current paradigm of participatory AI. Finally, we identify a set of empirically-grounded opportunities for future research to explore ways to improve the current state of stakeholder participation in AI design.


\section{Stage One: Establishing a Framework for Evaluating Participation}
\label{literature_participation}

\subsection{Related Work}

Given the heterogeneity of participatory AI, it remains challenging to have substantive discussions about the extent to which a particular participatory approach achieves one's goals (e.g., for inclusiveness or empowerment) \cite[e.g.,][]{arnstein1969ladder,fung2006varieties}.
To this end, the first goal of this paper is to develop a conceptual framework as a point of reference that can provide theoretical grounding for such evaluations.

Recent scholarship has started working toward typologies for classifying the different modes of participation in AI.
\citet{sloane2020participation} elucidates the distinction between participation as work, as consultation, and as justice; \citet{Birhane2022power} identifies three objectives of participatory AI: for algorithmic performance improvements, process improvements, and collective exploration; the Ada Lovelace Institute \cite{stewardship_2021} proposes a ``spectrum of participation'' to describe five levels of participatory mechanisms, from inform, consult, involve, collaborate, to empower; similarly, \citet{berditchevskaia2021participatory} conceptualizes four levels of participatory AI: consultation, contribution, collaboration, and co-creation.
While these are valuable starting points for conceptualizing and defining participatory AI, as \citet{Birhane2022power} points out, ``there are yet no clear consensus on what minimum set of standards or dimensions one should use to assess or evaluate a given potential participatory mechanism.''

\subsection{Participatory Traditions Across Disciplines}

To bring some structure to our analysis of the current heterogeneous state of participatory AI, we took a step back from the contemporary AI literature and broadly reviewed existing literature on participatory traditions from multiple disciplines, including: technology design \cite[e.g.,][]{simonsen2012routledge, muller1993participatory}, political theory \cite[e.g.,][]{fung2006varieties,arnstein1969ladder,polletta2012freedom}, and the social sciences \cite[e.g.,][]{unertl2016integrating}. 
We pulled from works that specifically addressed the nature of participation in policy and design processes, as well as works cited by more recent participatory AI papers (e.g., \cite{sen1986social}, as invoked by \cite{lee2019webuildai}). 
In addition to foundational theoretical and methodological works in these traditions, we included literature that offered critiques and challenges to participatory practices in their respective fields \cite[e.g.,][]{cooke2001participation}. 
The specific traditions we drew from for our synthesis include user-centered design, service design, participatory design, co-design, and value-sensitive design, as well as participatory action research, participatory democracy (including deliberation theory), social choice theory, and mechanism design. Below we provide a high-level overview of each of these traditions and their corresponding tenets.

\subsection{Approaches to Stakeholder Participation}
\label{prior-approaches}

\begin{itemize}[leftmargin=*]
    \item \textbf{User-centered design (UCD):} involving end users during need-finding and evaluation of designs.
    The double-diamond UCD process has become the most common design process in industry practice \cite{doublediamond}. Technology designers engage end users in identifying user needs and assessing design ideas \cite{mao2005state,olson2014ways}.\looseness=-1 

    \item \textbf{Service design:} involving direct and indirect stakeholders impacted by a technology service in its design. 
    Using methods such as service blueprints and value stream mapping, service designers consider how their design decisions impact stakeholders \cite{forlizzi2018moving, forlizzi2013promoting,forlizzi2018moving,saad2020service,holmlid2012participative}.\looseness=-1
    
    \item \textbf{Participatory design (PD):} incorporating diverse voices in design in order to challenge power structures. 
    In contrast with UCD and service design, PD is designed to enable non-expert stakeholders to provide direct input on technology design \cite{gregory2003scandinavian,simonsen2012routledge}.
    PD provides guidance for practitioners to 
    create a hybrid space (i.e., neither in the stakeholders or the designers' domain) that allows diverse non-expert stakeholders to meaningfully contribute,  
    while uncovering and challenging the power dynamics between stakeholders and designers, as well as within groups of stakeholders
    \cite{le_dantec_strangers_2015,harrington2019deconstructing,muller1993participatory}.\looseness=-1

    \item \textbf{Co-design:} creative cooperation between stakeholders and technology designers across the whole span of a design process \cite{def-codesign-sanders,co-design-as-joint-inquiry-imagination,def-codesign-Kleinsmann}. While closely related to PD, co-design typically lacks its explicitly political component \cite{co-design-as-joint-inquiry-imagination}. Further, co-design often focuses more explicitly on designing specific products or artifacts \cite{def-codesign-Kleinsmann}, while PD may focus more on the larger sociotechnical system in which particular artifacts will be embedded \cite{simonsen2012routledge}.
    
    \item \textbf{Value-sensitive design (VSD):} accounting for the values of direct and indirect stakeholders. VSD is similar to PD, but focuses on identifying and incorporating the values of direct and indirect stakeholders in technology design \cite{friedman1996value,friedman2002value}.\looseness=-1 

    \item \textbf{Participatory action research (PAR):} engaging stakeholders as co-inquirers to co-construct research plans and interventions \cite{elliot1991action, unertl2016integrating,wallerstein2006using,hayes2014knowing, rasmussen2004action}. Originating in the social sciences (and adopted in HCI \cite{hayes2014knowing, harrington2019deconstructing}), PAR involves stakeholders collaborating with researchers to develop research questions and study designs, conduct research with researchers, and interpret results \cite{unertl2016integrating,wallerstein2006using,hayes2014knowing,rasmussen2004action,elliot1991action}, focusing on research rather than design (as in PD) \cite{gleerup2019action}.\looseness=-1 
    \item \textbf{Social choice theory (SCT) and mechanism design:} quantitative aggregation of stakeholder preferences.
    SCT focuses on identifying people's preferences for (e.g.) public policies and mathematically aggregating those in a preference-ranking model \cite[e.g.,][]{arrow2012social,sen1977social,sen1986social, kahng2019statistical, saha2020measuring}. 
    Prior work in the field of mechanism design has used SCT to develop a framework for including stakeholders in algorithmic decision-making \cite{abebe2018mechanism,finocchiaro2021bridging,hitzig2020normative,viljoen2021design}.\looseness=-1
    
    \item \textbf{Participatory democracy and civic participation:} involving citizens and stakeholders in civic decision-making. Although sharing a focus on policy outcomes with SCT, participatory democracy has a broader focus on methods for democratic decision-making, political action, and public engagement \cite[e.g.,][]{arnstein1969ladder,fung2006varieties,polletta2012freedom, lippmann1993phantom, dewey19541927, macpherson_democratic_1973,polletta2012freedom}.\looseness=-1
    
    \item \textbf{Deliberation theory:}~qualitatively weighing and discussing competing perspectives and policies.
    Deliberation theory emerged in response to mechanistic approaches to aggregating stakeholder preferences (e.g., social choice theory) \cite{oberg2016deliberation,fishkin2005experimenting,owen2015survey}. It emphasizes the importance of bringing together small groups of people to discuss and qualitatively weigh competing arguments for policies \cite{fishkin2005experimenting}.\looseness=-1
    
\end{itemize}

For many of these traditions, stakeholders' power is manifested through the creation and deliberation of design alternatives that would not have been possible without their contribution \cite{arnstein1969ladder, bratteteig2016unpacking, bodker2018facing, bratteteig2012disentangling}. 
Yet, in practice, previous surveys of participation have observed that many participatory techniques fail to truly empower stakeholders in a way that positions them as co-creators or co-owners in the design process. 
That is, in practice, many participatory approaches may extract the input of stakeholders for goals not defined by or even necessarily shared with those stakeholders \cite{beck2002political, balka2010broadening, viljoen2021design, hitzig2020normative, robertson2020if, sloane2020participation, miceli2020between,cooke2001participation, arnstein1969ladder, cooke2001participation, palacin2020design}.
Additionally, critics have noted that researchers and practitioners have difficulty putting aside their own notions of what the design should be and what the intervention should look like, thus blocking possible new avenues for change originating from stakeholders' ideas \cite{blomberg2012ethnography}.
Design processes in which stakeholders are only selecting or evaluating predefined choices have been described as forms of ``pseudo-participation'' \cite{palacin2020design,robertson2020if} in which stakeholders lack agency in decision-making and agenda setting in the design process \cite{pedersen2016war, bodker2018facing, bratteteig2012disentangling}. 
Further, critiques have highlighted the risk of de-politicizing participation through methods that focus on individuals' preferences rather than collective goods or examination of power structures \cite{beck2002political,viljoen2021design,hitzig2020normative}.

To address this conundrum, participation scholars have called on researchers and practitioners to shift their role away from directly engaging in design activities towards configuring the context and procedures through which meaningful stakeholder participation and design contribution can take place \cite{vines2013configuring, def-codesign-sanders}.
To help enact this role shift, participation scholars have focused on the importance of reflexivity as a way to help researches and practitioners understand and beneficially redirect their relative power over stakeholders \cite{vines2013configuring, bodker2017tying, co-design-as-joint-inquiry-imagination, rigour-in-PD}. 
Nonetheless, given existing power structures within communities, and specifically between participants and designers, simply being reflexive may not be sufficient in centering the stakeholders' values in the design process. \cite{cooke2001participation,harrington2019deconstructing}.

\subsection{Parameters of Participation} 
\label{conceptual_framework}

\begin{figure*}[t]
    \captionsetup{width=1\linewidth}
    \includegraphics[width=0.8\textwidth]{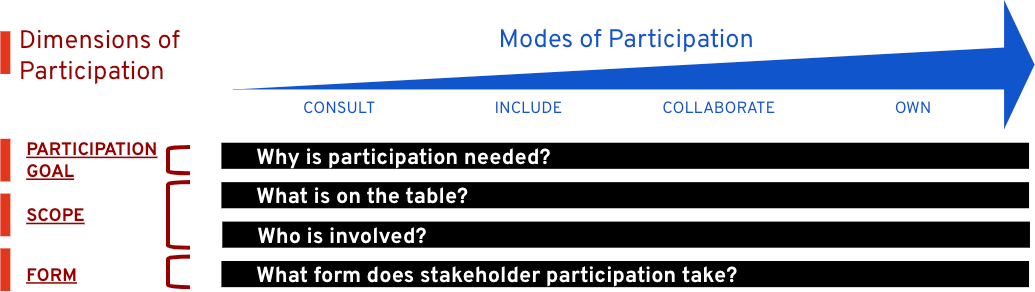}
  \caption{\framework: a framework derived from a synthesis of prior literature on stakeholder participation\looseness=-1} 
  \label{fig:CF}
\end{figure*}

We used affinity diagramming to identify common threads across these nine different traditions of participatory scholarship discussed in the previous section. We chose affinity diagramming because it is a common method from HCI used to make sense of an unstructured problem space \cite{holtzblatt1997contextual}.
Three of the authors iteratively engaged in affinity diagramming using a virtual whiteboard to make connections across the various schools of participatory thought and practice with a focus on isolating how each tradition framed the relationship between participation and agency, as well what key questions were articulated across traditions (even if they came to different answers). 
We iterated through several rounds of discussion to resolve disagreements in order to reach consensus on our final version of the framework. 
Figure \ref{fig:CF} illustrates the resulting \framework, which includes a spectrum of different modes of stakeholder participation (in blue), from consultative to stakeholders ``owning'' the project on the horizontal axis, and a set of key dimensions (in red) that inform the goals and methods of participation on the vertical axis.

\subsubsection{Dimensions of Participation}
\label{dimensions}


The dimensions of participation represented on the vertical axis comprise the \textit{goals} (i.e., why participation is needed), \textit{scope} (i.e., what is on the table; and who is involved) and \textit{methods} (i.e., the form of stakeholder participation) that shape participatory approaches to AI. We frame these dimensions as questions for researchers and practitioners to ask themselves throughout the lifecycle of their participatory intervention---or questions that may be asked by impacted stakeholders, participants in a participatory AI initiative, or policymakers or others looking to better understand participatory AI efforts. \looseness=-1 

\begin{itemize}[leftmargin=*]
    \item \textit{Why is participation needed?} Motivations to engage stakeholders broadly fall under two categories: The first is an \textit{instrumentalist} motivation that stakeholder participation will result in better outcomes \cite{bjerknes1995user}. However, ``better'' might mean a range of things such as an outcome that is more reflective of stakeholders' values, preferences, and needs \cite{polletta2012freedom,simonsen2012routledge}, or an outcome that leads to more profit or legitimacy for the designers \cite{cooke2001participation}--each of these possibly in tension with one another. The second is a \textit{normative} motivation: i.e., participation is important because inclusion of the people who may be impacted by the technology or policy being designed is simply the right thing to do \cite{polletta2012freedom}. 
    For example, proponents of participatory democracy argue that participation is normatively important, as it aligns organizers' ideological commitment to an egalitarian society to the methods adopted to achieve that society \cite{polletta2012freedom}. 
    Participation may be conducted with both goals in mind, as in deliberation theory, which aims to give people a voice in shaping more effective policy decisions that lead to political equity \cite{owen2015survey,fishkin2005experimenting}.\looseness=-1
    
    \item \textit{What is on the table?} The extent to which stakeholders are able to determine which design decisions they will inform or even make is a key differentiator across the various participatory schools of thought we surveyed \cite{arnstein1969ladder,fung2006varieties}. For instance, in SCT, what is made available to participants is a set of of policy alternatives for them to rank. 
    Rather than having participants propose or generate such alternatives, this set of options is often determined a priori by technical experts. 
    It is important to note that this approach assumes that participants are able to express their desires in terms of ranking pre-defined alternatives (which may not be the case should their preferences fall outside of the scope of those options which they are presented) \cite{robertson2020if,finocchiaro2021bridging,viljoen2021design,hitzig2020normative}. 
    By contrast, in PD, PAR, and other participatory approaches in HCI, identifying `what is on the table' is often the core focus of the participatory approach. This can range from stakeholders contributing to specific design decisions about how a technology will be built, 
    to more broadly designing the social or sociotechnical system in which those technologies will be embedded \cite{simonsen2012routledge,muller1993participatory,forlizzi2013promoting}, to even designing the research project as a whole, as in PAR \cite{unertl2016integrating,wallerstein2006using,hayes2014knowing,rasmussen2004action, harrington2019deconstructing,dearden2014scaling,rigour-in-PD}.\looseness=-1
    
    \item \textit{Who is involved?} The rationale and selection criteria for who gets to participate substantially differs based on the participatory tradition. For instance, service design and PD involve both direct and indirect stakeholders (i.e., beyond the end user) as well as involve participants with less technological expertise who otherwise might be excluded from the design of complex systems \cite{forlizzi2013promoting,forlizzi2018moving,simonsen2012routledge,muller1993participatory, light2008designing}. 
    PD strategically focuses on engaging participants who will be affected by design decisions, often intentionally involving marginalized voices specifically, rather than seeking to be broadly inclusive of all possible stakeholders \cite{simonsen2012routledge,light2008designing}. 
    In participatory democracy, some approaches involve participation of members of the general public, while others involve a cadre of expert administrators or representatives acting (nominally) on the community's behalf \cite{fung2006varieties,arnstein1969ladder}. In deliberation theory, a group of participants is selected by a randomized sample in order to nominally represent the range of views held by the population at large \cite{oberg2016deliberation,fishkin2005experimenting}. SCT involves a larger number of participants, typically selected by the team of technical experts developing the models \cite{hitzig2020normative,viljoen2021design}.
    
    \item \textit{What form does participation take?} In SCT, people participate by expressing preferences for policies, often by ranking alternatives on polls \cite{robertson2020if,hitzig2020normative}. However, deliberation theory argues that such approaches fail to account for how people learn new information and update their beliefs over time \cite{fishkin2005experimenting,fishkin2017prospects}. Instead, in deliberation theory, participation involves a dialogic approach---bringing stakeholders and experts together to discuss trade-offs of policies and re-evaluate their opinions in light of new evidence \cite{oberg2016deliberation,fishkin2005experimenting}. In PD, participants are brought together for mutual learning with designers in ``hybrid spaces'' such as workshops involving (e.g.) storytelling, role-play, and co-creation of prototypes \cite{muller2002participatory,simonsen2012routledge}.\looseness=-1 

\end{itemize}

\subsubsection{Modes of Participation: From Consulting to Ownership}
\label{modes}
How best to maneuver as a practitioner or researcher in varying modes across these dimensions is a motivating dilemma within various branches of participatory scholarship. For example, in participatory democracy, questions of power dynamics include how policy design processes might be shared democratically \cite[e.g.,][]{polletta2012freedom} and how such design processes might avoid being tokenistic \cite[cf.][]{arnstein1969ladder} or reifying existing power dynamics among groups of stakeholders \cite[cf.][]{cooke2001participation}. In PD, with its origin in workplace labor organizing, participation is explicitly intended to equalize power relations between stakeholders and designers (as well as among different groups of stakeholders) \cite{simonsen2012routledge,muller1993participatory,muller2002participatory,muller2012participatory}. 
Crucially, in many traditional participatory scholarship and critique surveyed for this article, the goal of full stakeholder empowerment is achieved in practice only in situations where stakeholders have the ability to make decisions about both the \textit{design} of systems or policies, as well as the \textit{process} by which such a system or policy is developed \cite[e.g.,][]{polletta2012freedom,cooke2001participation}. \looseness=-1

Reflecting the central role that stakeholder decision-making agency traditionally has in theorizing and evaluating participatory practice, we include what we call four \textit{modes of participation} as part of the \framework. This move follows substantial prior scholarship that has articulated varying degrees of participatory power, such as Arnstein's ladder of civic participation \cite{arnstein1969ladder}, Fung's democracy cube \cite{fung2006varieties}, and others \cite[e.g.,][]{algo_2021}. 
These four modes of participation form a spectrum. 
There are participatory approaches that \textit{consult} with stakeholders (e.g., expressing preferences for policies in a ranking); those that more substantively \textit{include} stakeholders (e.g., deliberating about pre-defined policy options); and those that \textit{collaborate} with stakeholders (e.g., having stakeholders co-create candidate designs for selection). 
The most substantive form of participation occurs when stakeholders \textit{own} the design process, playing a central role in shaping the procedures for deliberation as well as deciding on the outcomes of the process (e.g., co-constructing research questions and methods, or the ``citizen control'' described by \citet{arnstein1969ladder}). 
We combine these four \textit{modes of participation} with the four \textit{dimensions of participation} previously discussed above to derive our conceptual framework of participation (Figure \ref{fig:CF}), which we leverage for our analysis of participatory approaches to AI design in this paper (Figure \ref{fig:CAM}).\looseness=-1 

It is important to note here that this model is meant to provide at its base a descriptive schema through which we can map participatory AI interventions according to a shared set of parameters that carry across a wide variety of contexts. This schema is not meant to necessarily imply that transferring ownership to stakeholders is the appropriate goal for any project dubbed participatory (even if some of the scholarship and critique reviewed would make this normative claim). On the contrary, our intervention here---while drawing much from existing traditions of participation---is motivated by the desire to articulate the terrain between transactional preference elicitation on one hand, and transformative subversion of power dynamics on the other. Moreover, our goal here is not to assign an evaluation purely at the level of which \textit{mode} of participation is enacted on any given project, but rather to provide a framework to understand how the particular configuration of participatory \textit{dimensions} shape the agency of the stakeholders involved.

\section{Stage Two: Examining Current Participatory Approaches to AI Design}

\subsection{Methods}
\label{methods}
We applied our \framework~(figure \ref{fig:CF}) framework to analyze participatory AI as an emerging practice, including the 
extent to which the participatory methods used in AI design are aligned with their stated goals for participation (i.e., the degree to which they engender stakeholder agency). 
The data we used to inform our analysis includes a corpus of 80 research articles in which authors report making use of participatory methods (see section~\ref{corpus_development}) and 12 semi-structured interviews with researchers and practitioners who were authors of at least one of the papers in our corpus (see section~\ref{interviews}).\looseness=-1 

\subsubsection{Corpus Development and Analysis}
\label{corpus_development}
To develop our corpus, we compiled a set of research articles that reported use of participatory methods in their AI design approach. 
We searched for relevant research articles within conference proceedings accessible from the ACM Digital Library, as well as premier AI venues, such as the Conference on Neural Information Processing Systems (NeurIPS), the International Conference on Machine Learning (ICML), and the Association for the Advancement of Artificial Intelligence (AAAI). 

Our first pass in identifying articles for inclusion was based on results from the following search query across all available text fields: ``participatory design'' OR ``PD'' OR ``co-design'' \textit{AND} ``artificial intelligence'' OR ``AI'' OR ``machine learning'' OR ``ML.'' 
Informed by an initial round of data analysis, we also conducted a second pass search to broaden the scope of our corpus. We conducted the search on the same databases but with the following broader search queries derived from key terms in papers from our first pass: ``participation'' or ``participatory'' or ``co-design'' or ``value-sensitive'' or ``human-centered'' or ``deliberation'' and ``artificial intelligence'' or ``AI'' or ``machine learning'' or ``ML'' or ``algorithm'' or ``algorithmic.''\looseness=-1

We reviewed the query results to identify a final set of relevant articles in which (1) the research described the design of some type of AI system broadly construed (e.g., autonomous robots, natural language processing, or machine learning), (2) the research described design efforts leveraging participatory techniques, rather than, for instance, calling for (or critiquing) participation in AI \cite[e.g.,][]{loi2018pd,cscw2018wkshp,sloane2020participation}. Based on these inclusion criteria, we identified $80$ relevant articles for inclusion into our corpus.  A full listing of the relevant articles can be found in Appendix \ref{appendix:corpus}. The resulting set of articles in this corpus reflects research across a wide range of application domains (e.g., education, healthcare, public services, content moderation), a diverse set of AI/ML technologies (e.g., predictive models, recommendation systems, classification systems, human-robot systems), and a broad range of organizational affiliations (e.g., academia, industry, public sector, nonprofit organizations). 

Taking a “descriptive analytical” approach~\cite{arksey2005scoping}, we used the logic and content of the \framework~to code the corpus of research articles, using the \textit{dimensions of participation} and \textit{modes of participation} as codes.
Two authors iteratively coded the articles in the corpus using these codes, resolving any disagreements about codes by discussion in order to reach consensus. \looseness=-1

\subsubsection{Researcher and Practitioner Interviews.}
\label{interviews}

\begin{table*}[]
    \centering
    \begin{tabular}{c c c c} 
        \toprule
        \textbf{Identifier} & \textbf{Sector} & \textbf{Application Domain} & \textbf{AI Task} \\[0.5ex] 
        \midrule
         P01 & Academia & Digital Humanities & Classification \\ 
         P02 & Non-Profit & Social Computing/Crowdsourcing Platforms & Prediction \\
         P03 & Academia & Social Computing/Crowdsourcing Platforms & Prediction \\
         P04 & Academia & Astronomy and Healthcare & Prediction \\
         P05 & Academia & Social Welfare & Ranking \\
         P06 & Industry & Education & Conversational AI\\
         P07 & Industry & Financial Services & Prediction \\
         P08 & Academia & Social Welfare & Prediction \\
         P09 & Academia & Social Welfare & Prediction  \\
         P10 & Industry & Arts \& Entertainment & Prediction\\
         P11 & Academia & Online Moderation, Gig Work & Prediction \\
         P12 & Public Sector & Healthcare & Prediction \\ 
        \bottomrule
    \end{tabular}
    \vspace{0.1cm}
    \captionsetup{width=.75\textwidth}
    \caption{\textbf{Interview Study Participants}}
    \label{tab:participants}
\end{table*}

In order to understand the motivations, processes, and challenges that researchers and practitioners face in pursuing participatory AI work, we conducted IRB-approved, semi-structured interviews with 12 researchers and practitioners who have worked on an AI project leveraging participatory methods, and who were authors on papers included in our corpus. We recruited 23 interviewees based on a purposive sample~\cite{palinkas2015purposeful} of authors of articles in the corpus, of whom 12 agreed to participate.
We recruited these 23 interviewees based on their varied experience in leveraging stakeholder participation in AI design on projects across a broad range of application domains, and to cover a wide range of the disciplinary backgrounds and sectors prevalent in the sample of research literature corpus. 
See Table~\ref{tab:participants} for more detail about the interviewees, with some details abstracted to preserve interviewees' anonymity.\looseness=-1

Each interview lasted 45-60 minutes, and all were conducted remotely on a video conferencing platform due to the COVID-19 pandemic. All interviews were recorded, transcribed, and anonymized, omitting specific details of projects as needed.
The interviews centered around questions related to the project described in the article in our corpus that they co-authored. This included questions about what motivated them to include stakeholders on their project, how they determined which stakeholders were involved, what role stakeholders played on the project, and what challenges they faced in adopting a participatory approach to their AI design work. We then asked interviewees to describe the actual participatory process they undertook in their work, as well as their aspirational vision for what they thought the participatory process \textit{should} look like based on what they learned from their previous experiences.\looseness=-1 

We took a hybrid approach \cite[e.g.,][]{fereday2006demonstrating} to analyzing our interview data, using both a codebook---based on our dimensions and modes of participation from the framework---and inductively through inductive thematic analysis~\cite{braun2006using,braun2019reflecting} (see~\cite{chouhan2019co, jiang2019recent} for precedents of this hybrid approach to qualitative data analysis). 
This hybrid approach allowed us to combine theories of participation from the framework (i.e., our codebook) and empirical evidence (i.e., from inductive thematic analysis) to better understand current approaches to stakeholder participation in AI design and development.
First, we analyzed the interview data using primarily the dimensions from our conceptual framework (Figure \ref{fig:CF}).
Then, after extracting excerpts of the transcripts based on the dimensions, we used inductive thematic analysis \cite{braun2006using} to identify themes from these excerpts.
Using a virtual whiteboard to cluster the excerpts, all authors iteratively clustered the excerpts into themes, collectively discussing the emerging themes and revising our clusters during the discussions to resolve any disagreements about codes \cite{braun2006using,olson2014ways}.
Examples of such themes include ``anxieties around how to validate participatory approaches,'' ``proxy-based participation,'' and ``consultancy mode of collaboration.''
This approach to inductive thematic analysis, much like affinity diagramming, is widely used in qualitative-interpretive research in HCI and the social sciences \cite{olson2014ways}.

\subsection{Findings: Current Participatory AI Landscape - A Largely Consultative Terrain}
\label{findings-landscape}

\begin{figure*}[t]
    \captionsetup{width=1\linewidth}
    \includegraphics[width=1\textwidth]{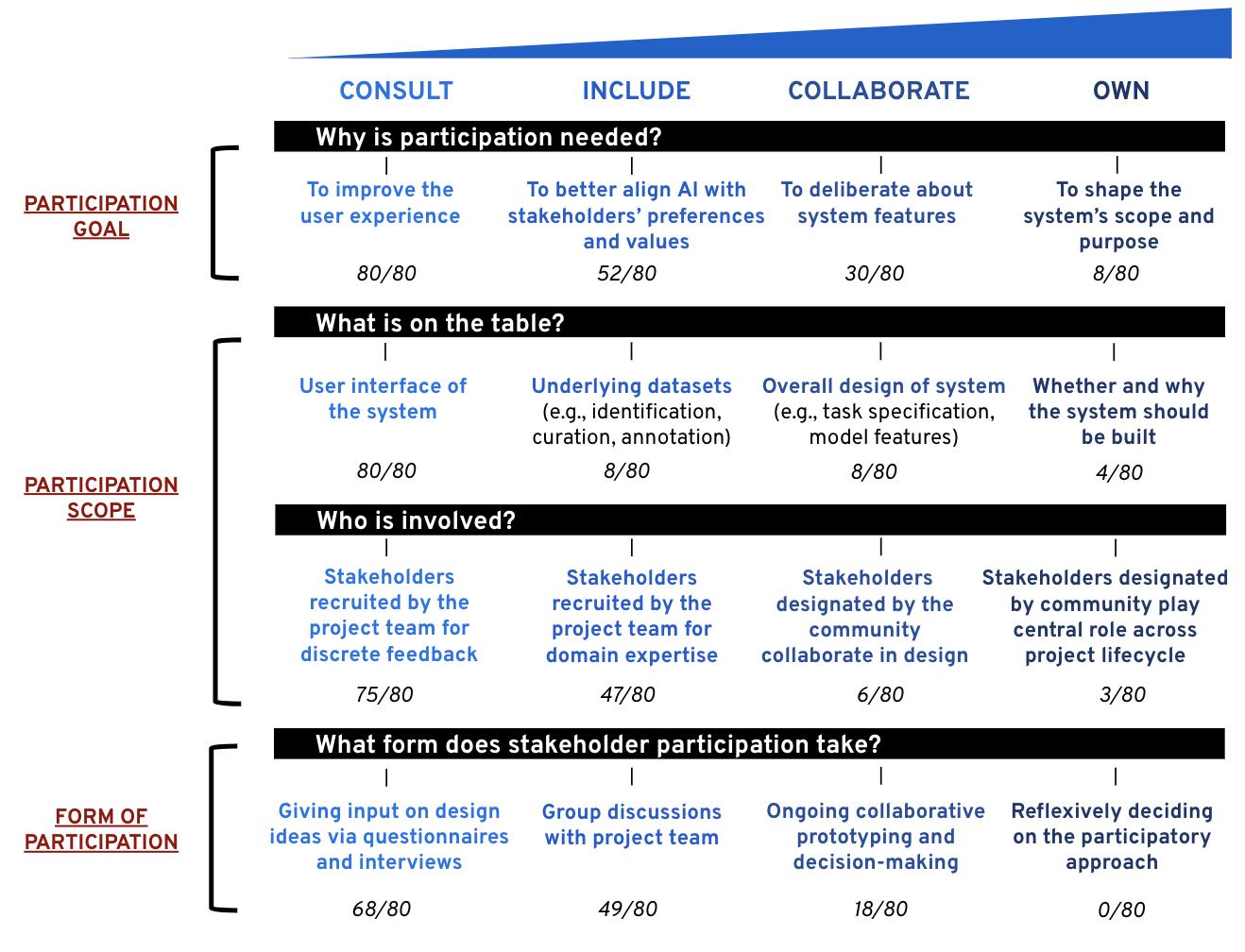}
  \caption{Participatory AI Projects Mapped to Conceptual Framework} 
  \medskip
  \small{Results from coding the $80$ research articles in our corpus according to our conceptual framework (see section~\ref{conceptual_framework}). This figure is annotated with text specifying exemplar approaches pulled from both the corpus analysis and supporting literature review.}
  \label{fig:CAM}
\end{figure*}

\subsubsection{The majority of projects operationalized stakeholder participation via discrete preference or value elicitation}

Our analysis of the corpus of research articles indicates that the current state of participatory AI design falls largely under the consultative mode in which preferences and values are elicited from stakeholders through a variety of techniques. The output of these elicitation processes is then analyzed by AI researchers in a broader loop of design and development activities. 

From our corpus analysis, we found that in motivating ``\textit{why is participation needed}'', the majority of projects described their goal as \textit{to improve the user experience} (80 out of 80) or \textit{to better align AI with stakeholders' preferences and values} (52 out of 80). This reflects a state of participatory practice which focuses less on collaboration and stakeholder ownership and more on consultation and inclusion. In fact, only 10\% (8 out of 80) of the projects involved stakeholders with the goal of \textit{shaping the system's scope and purpose}. See Figure~\ref{fig:CAM}, first row.

In terms of the scope of participation, the majority of projects surveyed focused their participatory interventions exclusively on informing selected aspects of the user interface, thus restricting ``\textit{what is on the table?}'' to a particular component of the system in question (and notably one that does not directly link to the algorithmic or machine learning aspects of the system). 
In fact, \textit{all} projects involved stakeholders in shaping the user interface (UI) design of the AI system, such as designing the explanations around AI outputs. 
Shaping the design of AI models is rarely on the table for stakeholders. Only 10\% (8 of 80) of the projects allowed stakeholders to provide input into the type of model and features used, the design of objective or loss functions, or (for AI classification models) the decision thresholds.
None of the projects allow stakeholders to rule out AI as a solution, since these projects were defined as ``AI projects'' in the first place. Mapping these observations onto the levels of participation framework, we can see that a vast majority of projects fall into the consultative realm in terms of ``\textit{what's on the table}``. See Figure~\ref{fig:CAM}, second row.

As it relates to ``\textit{who is involved?},'' the majority of stakeholders brought onto the projects in our corpus were identified and chosen by project leads (74 of 80).
Often (47 of 80), project leads' choices were based upon what they presumed to be the background of that group/person, recruiting stakeholders either opportunistically (``\textit{just talk to some of your friends}'' [P10]) or through an opt-in mechanism (``\textit{Some people just kind of came to (our) team and said, hey, can we work with you?}'' [P9].
This contrasts with the much smaller number of projects that leveraged a community-engaged approach in establishing relationships with stakeholder groups and leveraging those relationships to help designate participants for design activities (6 out of 80).
Even fewer projects (3 out of 80) had community stakeholders involved throughout the life cycle of the AI design project. 
Overall, reliance on project leads' criteria for deciding ``\textit{who's involved}'' across the  majority of projects means less agency is given to community stakeholders in terms of design decision-making. Further, the relative scarcity of relationship-building and stakeholder involvement throughout the end-to-end AI lifecycle points more toward an ad-hoc mode of participation of current practice. See Figure~\ref{fig:CAM}, third row.

As it relates to participatory methods, 85\% (68 out of 80) of the projects engaged stakeholders in eliciting preferences or values through methods such as surveys and interviews, role-playing, and collaborative story-boarding. 
The majority of projects (57 out of 68) engaged with stakeholders only once, meaning that these high-level inputs were the only inputs stakeholders provided.
For such projects, participation primarily took place during designated windows once the system's scope and purpose were already determined---or were specifically scoped to aspects of the user interface, rather than the underlying AI model, as previously described. 

In fact, fewer than a quarter (18 out of 80) of the research articles we reviewed described engaging with the same participants more than once throughout the research project. As an example, in their research article on participatory design of social robots, \citet{lee2017steps} adopted a six-step approach to their participatory design process. This approach involved interviewing the participants to learn from their perspectives, brainstorming design ideas with participants, and validating final design ideas with the community. This ``multi-phase'' approach \cite[cf.][]{georgiou2020applying} affords stakeholders more opportunities for participation throughout the design process than single instances. That said, none of the 80 research articles we analyzed involved stakeholders early on in reflective deliberations regarding the participatory design process or their proposed role in it. See Figure~\ref{fig:CAM}, fourth row.

\subsubsection{Researchers and practitioners feel they have to substantially scale back their participatory ambitions} 
\label{constraints}

Interviewees across industry, the public sector, and academia reported that top-down organizational constraints limited the time and resources available for participatory approaches. This narrowly limited the dimensions of ``\textit{what is on the table}'' and ``\textit{which stakeholders are involved}'' for their projects. 
Often, only \textit{how}---not  \textit{whether}---AI will be deployed was subject to discussion with stakeholders.
Corporations', research teams', and government agencies' priorities, mandates, and resources were often reported to be in tension with interviewees' desire to empower stakeholders. 
Multiple interviewees in the public sector, for example, described how legislative rules placed strong constraints regarding who could or could not be involved: ``\textit{children on the ground [...] They have no agency in this. They're also stakeholders, but they don't have any agency, because by law they’re not capable of making decisions}'' [P9]. 

For industry researchers, we heard how participatory projects needed to be scoped to align with formal management indicators and metrics: ``\textit{OKRs [Objectives and Key Results], KPIs [Key Performance Indicators] and all this sort of stuff... those come from the top of the organization, and then trickle down}'' [P2]. This interviewee felt that they needed to ``\textit{fight more with management}'' [P2] to advocate for stakeholders' interests to elevate them to the level of organizational priorities. This uphill battle in securing stakeholder-based priorities led to researchers and practitioners making compromises regarding the amount and degree of stakeholder participation. The researcher or practitioner in such situations was often put in the situation where the best they could do was talk ``\textit{to a bunch of people}''---yet without truly empowering them---``\textit{we call it `human-centered design'}'' [P9].


When there is corporate interest in participatory projects, AI researchers and practitioners faced unanticipated expectations to include stakeholders other than those they were initially focused on based on ``\textit{the business needs for that particular AI system}'' [P7] or to include the sponsors who contributed to funding the project [P6]. AI researchers and practitioners working on public sector projects also often face a what seems like an ever-growing set of stakeholders across various groups spanning organizational boundaries: 

\begin{quote}
    \textit{And so if you think about stakeholders [...] you have agencies, you have the department at the state level, but you also have the court, [...] you have judges, you have attorneys, but when you want to add algorithms into the mix, you also have to talk about developers or, you know, the various IT companies that these people contract with, right? [...] that's the world that they live in, and it's a very messy world.} [P9]
\end{quote}

For academic researchers conducting participatory work, the long-term relationship development needed for adequate understanding of stakeholders' needs and social contexts is further undermined by the demand for rapid publication of results, the need to secure funding for students, and tenure requirements for faculty \cite[cf.][]{black2020call}. One participant described how they ``\textit{set up a multi-year project in the hope that I would get external funding down the line, which would fund this project}'' [P9], but this hope for funding is not a sustainable bet for many, and can often leave stakeholders in the lurch if this funding runs out and the engagement ends prematurely \cite[cf.][]{harrington2019deconstructing}. Long-term relationship development is also not well-aligned with the discrete semester-long engagements that many academic research projects are scoped around [P3]. 

Across all sectors, organizational timelines and resources were described as being at odds with involving a wide range of stakeholders, and certainly ill-equipped to support the ``\textit{deep ethnography}'' [P9] that enables a rich understanding of the social landscape in which algorithms are being designed for.
Relatedly, interviewees described feeling caught between an idealized world of stakeholder empowerment and the practical constraints of available time and resources for participation. Interviewees described the desire for substantive stakeholder participation as an ``\textit{ambitious}'' [P10] but (perhaps unrealistically) idealistic goal, because ``\textit{we don't live in a perfect world}'' [P9]. 

One interviewee went further to describe the notion of full participation as an impossibility, as for them this would require participants ``\textit{literally coming to the office with me and making every decision with me and doing all these things; all of a sudden, they don't have a life to live, right?}'' [P10]. 
This straw-man description of participation seems to serve at one level to justify a diluted version of stakeholder involvement, and another as a source of anxiety about never being able to do participation in design any justice. 
Others were explicit that a more ideal form of participation involved the ``\textit{actual ability to change actual power and actual ability to choose a product direction and move it}'' [P3]. However, this interviewee went on to tell us that they felt ``\textit{a little jaded about companies (not) being willing to really change course}'' [P3] based on participants' desires. 

Interviewees' anxiety about not being able to achieve idealized modes of participation was occasionally expressed in feelings of guilt over not doing enough to engage more stakeholders in more ways. 
As one researcher told us, ``\textit{having more people involved is always better. And I think one thing that we didn't do very well is like expand past what was like logistically kind of easy for us [...] we just decided to start with the people we could easily contact}'' [P5]. 
However, as this interviewee went on to acknowledge, those stakeholder groups who were left out were ``\textit{the end goal of who we want to like benefit, right? And I think not having them in the picture is kind of bad}'' [P5]. 


In addition to tradeoffs about who was involved in participating, interviewees described how the pragmatic realities of participants' circumstances motivated them to adopt particular approaches to participation. For instance, much like the concern of P10 that participants ``\textit{have a life to live}'' and thus cannot provide input on every possible design decision, another interviewee shared how the approach they took to participation involved selected participants training models to make decisions in their stead (see section~\ref{proxies_participation} for more about this type of approach). For them, this approach was:

\begin{quote}
    \textit{...a way of like realizing this [constraint] in an actionable way. Because you can't have people constantly voting like 100 times a day over like where they think [resources should be allocated]. And you probably only have like a minute or a minute and a half to like actually [allocate each resource], 'cause you have to like coordinate a bunch of volunteers... And it's just very like operationally challenging.} [P5]
\end{quote}

Some interviewees [P9, P10, P11] justified such pragmatic choices by arguing that participation---any form of participation, however limited
---would be beneficial for stakeholders and better than the alternative of not having participated at all.
As one researcher told us: ``\textit{the reality is a lot of times is, when I say participatory, I mean two hours [...] but I would argue it's better than nothing or like no participation}'' [P10]. As they elaborated, ``\textit{I'm so in the mindset that if you get 80\% closer to something, that's a win. If you genuinely get closer to something, I think that's better}'' [P10], while another interviewee expressed the importance of doing ``\textit{whatever I can, and it won't be perfect, and it will be rough and it will be messy}'' [P9]. Meanwhile, others highlighted the importance of developing ``\textit{lighter-weight methods}'' [P3] such as surveys to follow up with stakeholders who were less interested in ongoing engagement via traditional PD style workshop sessions.

\subsection{Findings: Proxy-Based Participation}
\label{findings-proxies}

Across our corpus and interview data, we found that instead of directly engaging the affected stakeholders, AI system designers have come to embrace a proxy-based approach to participation. 
In particular, we identified three proxy-based tactics that AI system designers adopted to ostensibly incorporate stakeholders' voices into the AI design process: (a) leveraging people who work with or are otherwise familiar with stakeholder communities to serve as stand-ins, (b) the introduction of User Experience (UX) or Human-Computer Interaction (HCI) practitioners as mediators that sit between AI experts and stakeholders, and (c) the use of \textit{algorithmic proxies} to elicit and model stakeholders' preferences by training models.

\subsubsection{Stand-ins for affected stakeholders.~}
\label{proxies_participation}

For example, in some cases, participants involved in design were not necessarily members of an affected stakeholder group themselves, but rather individuals who the project team perceived as being able to stand-in and voice stakeholders' preferences and values based on lived and/or work experience \cite{luhtala2018proactive, metatla2019voice, candello2020co, smith2020keeping, yu2020keeping, cheng2021soliciting, neto_community_2021, ammitzboll_flugge_street-level_2021, candello2021museum, yadav2019breastfeeding}. 
Some researchers we interviewed relied on policymakers they believed could ``\textit{somehow summarize whatever the affected users are saying}'' [P7]. Others, such as one educational research project, described using stand-ins (e.g., educators) to role-play as other stakeholders (e.g., children) to help design human-robot interaction patterns [P6].
In their project on enhancing rescue work in the Finnish rescue department, \citet{luhtala2018proactive} asked workshop participants to ``consider different viewpoints i.e. customer, communication or rescue worker, and especially what would a good proactive rescue service based on a partnership between different stakeholders be like.'' In another example involving a health counseling bot, \citet{kearns2020wizard} leveraged actors trained to role-play as patients to explore the possible needs of caregivers.
    
Another type of representative stand-in took the form of recruiting participants with similar demographic backgrounds to a target stakeholder group of interest. One interviewee described recruiting ``\textit{people in your company who can sometimes represent the group that you're looking at}'' [P10]. Despite acknowledging that employees of a tech company might be ``\textit{strange in different ways}'' (i.e., different from other members of those demographic groups), they went on to describe how those employees can still provide AI designers with ``\textit{the stuff that's just kind of obvious if you grew up like [they] grew up, if you had the lived experiences that [they] have, when [they] use these kinds of products}'' [P10]. 

\subsubsection{UX/HCI practitioners as mediators between AI experts and domain stakeholders}
\label{design_expertise}
    
Another tactic described in the corpus and interviews consisted of team members described as having UX/HCI expertise ``\textit{function as mediators}'' [P4] between impacted stakeholders and AI researchers in a way that (perhaps unintentionally) limited the interaction between these groups.
UX/HCI researchers in this view are seen as experts in the participation itself and thus positioned to play a central role in scoping, designing, and managing participatory interventions. This work included deciding which stakeholders (or others) would be brought in as participants, what role those participants played, and how participants interacted with each other and other team members. 

Multiple interviewees [P2, P3, P7, P8, P9] and research articles in our corpus \cite[e.g.,][]{holten2020shifting} specifically invoked UX/HCI expertise as imbued with a 
capacity that allowed UX or HCI experts to act as a bridge between stakeholders on one side, and algorithmic researchers on the other side.
They were viewed as mediators that could ``\textit{speak the same language}'' and have a ``\textit{two-way street}'' conversation across disciplinary lines [P5].
Interviewees described a ``\textit{two-sided market}’’ [P10] between team members with AI expertise and those with UX/HCI expertise in which the former relied on the latter to ``\textit{equip the non-experts [i.e., participants] to be able to have the conversation}'' [P10].

This was especially important as stakeholders were described as ``\textit{bad at coming up with mental models of what they value}'' [P5], having no ``\textit{idea of how the underlying technology works}'' [P4], or otherwise incapable of understanding ``\textit{the impacts of different design choices}'' [P11].
In fact, enabling non-technical stakeholders to understand AI's capabilities and limitations repeatedly came up as a key obstacle to address to ensure meaningful participation: 

\begin{quote}
	\textit{You need to explain the impacts of different design choices, so that people can start the conversation. Otherwise, these non-[AI]-expert community members, they just cannot discuss it. They do not know how to discuss these issues, right? [...You need to] actually explain this complexity, output and potential outcomes to the community, help the communities actually figure out and then navigate this complex, complicated outcomes, and then identify the actually the solutions that works best for the community.} [P11]
\end{quote}

Some papers went further to argue that the mediating role played by UX/HCI resources also ``\textit{temper[ed] the technological possibilities of algorithms and highlight[ed] the concerns for practitioners, whose work practices are being renegotiated as part of an overall change}'' \cite{holten2020shifting}. 
As such, this intermediary role was justified not only as a way to mitigate the perceived deficits in stakeholders' technical literacy, but also as a way to offset the techno-optimism or techno-solutionism of algorithm designers. 
Yet, this empowerment of those with UX/HCI expertise came at the expense of stakeholders' positioning as well as the quality and quantity of their engagement with the larger project team. 
For team members more focused on algorithmic design, the central positioning of UX/HCI experts pushed their interaction with the stakeholders to the periphery: ``\textit{HCI people talked to [the participants] a lot, but I hardly got to talk to them at all}'' [P5]. 
    

\subsubsection{Algorithmic proxies for participation.~}

In addition to these human proxies, other projects we analyzed from our corpus \cite[e.g.,][]{lee2019webuildai, cheng2021soliciting, zhang2020joint} drew on algorithmic proxies to ostensibly reflect stakeholders' interests. 
Some projects did so by directly soliciting stakeholder preferences via the AI systems they were developing. In these types of projects, project teams would ask stakeholders to train machine learning models, which would vote on binary outcomes (e.g., in a decision support tool) that nominally reflected stakeholders' preferences \cite[e.g.,][]{lee2019webuildai}. 
The assumption underlying this approach (informed by social choice theory \cite{kahng2019statistical, saha2020measuring}) holds that a predictive model trained on people's preferences is functionally equivalent to those preferences. 

Although stakeholders were involved in some fashion by training a model, this approach to participation was limited to providing training data or intervening in model training at one particular point in time. 
In this case, the set of preferences elicited from the stakeholder was only a \textit{snapshot} of the stakeholders' values at a particular time.
And, even when project teams aimed to repeatedly engage stakeholders in model training, it remained challenging to produce a model that is representative of the stakeholders' value systems given the constant flux of values and preferences that may exist for stakeholders at any particular point in time. 

In P5's case of developing an AI system for resource allocation, they reported that by the end of the elicitation process, many of their stakeholders prioritized values were very different from what they prioritized at the beginning of the process. For example, a stakeholder might begin the process by solely focusing on their own preference to access resources in their vicinity. However, once they started the process of pairwise comparisons, they might begin wondering if they should care about other factors such as poverty rate and income, which they realized might be the priorities of other people.
Looking back, P5 felt that designing a model that captures people's constantly evolving values was akin to ``\textit{trying to hit a moving target}.``


While many interviewees lauded such an iterative process through which stakeholders could attend to others' values to consider ``\textit{what's best for everyone}'' [P11], such dynamic modeling of stakeholders preferences becomes difficult to scale over time. 
Let us turn to case of P11 in which they were developing a matching system on a collaboration platform.
P11 explained that while there are ``\textit{hundreds of thousands of members}'' on the platform, ``\textit{there are only a few hundred members}'' that are available for matching at a given time.
The ever-changing composition of available members means that the definition of the \textit{best} matching algorithm is constantly changing.
Given the complexity of value elicitation at this scale, instead of attempting to directly model the myriad of preferences given the varying composition of available members, P11 created simulation environments in which they calculated tradeoffs that were similar to the ones the members were trying to navigate in real-time.
In these simulations, stakeholders preferences were mined but then further simplified to make the matching exercise more computationally tractable.

\section{Discussion}
\label{discussion}

Through our examination of participatory AI projects, we find that researchers and practitioners have difficulty mapping their participatory ambitions to the practical constraints they face on the ground.
As such, we find that the current state of participatory AI largely involves consultative approaches that elicit preferences and values from stakeholders. 
While there is certainly potential value to be gained in aggregating preferences across affected stakeholders, it is important to take note that only a handful of outlier projects surveyed in this study allowed stakeholders to drive or own any part of the design process itself.
As such, the current state of participatory AI leans heavily on a model where stakeholder engagement is highly discrete and programmed, affording little design agency to the stakeholders themselves. 

In our analysis, we also identified a growing trend in the use of proxy-based tactics in participatory AI including the use of representative stand-ins, the leveraging of UX/HCI practitioners as mediators between AI experts and stakeholders, as well as the use of algorithmic proxies to elicit or model stakeholders' preferences. While the first two examples of proxies are not necessarily unique to participatory AI, the latter is of particular interest in grappling specifically with whether, and if so how, AI presents a substantively new terrain for participatory interventions. Given the central role played by machine learning models in automating human decision-making and generating simulated user content (especially with the advent of large language models and generative AI) \cite{bommasani2021opportunities}, it is important to reflect on whether or how to best leverage these tools as stand-ins for stakeholders in participatory AI. 

Drawing from the above empirical findings and insights, we first focus our discussion below on the importance of aligning participatory methods to specific underlying participatory goals. In so doing, we call for a departure from conceptualizing participation as an all-or-nothing characteristic \cite[cf.][]{bratteteig2016unpacking} and advocate for identifying actionable goals that specify the level of agency and weight in decision-making stakeholders will be provided for any particular project (understanding that there is a spectrum of agency and decision-making authority, and that there is not one right solution for all projects). Second, we identify characteristics of AI systems (opacity and scale) that lead to particular challenges to meaningful participation. Third, we raise questions about the use of human and algorithmic proxies for participation.

\subsection{Moving beyond idealized aspirations toward specific outcomes}
\label{implications}

Instead of focusing on the idealized aspirations of participatory approaches, we recommend conceptualizing participation as a set of dimensions (see section~\ref{dimensions}) that need to be explicitly specified at the project level. This means moving away from conceptualizing participation as an all-or-nothing characteristic of a project \cite[cf.][]{bratteteig2016unpacking}, and setting specific goals regarding the extent to which stakeholders are able to contribute to the project. This also moves us away from a ``\textit{something is better than nothing}'' perspective, and forcing a specific discussion of how much agency and decision-making authority is meant to be provided to affected stakeholders. To aid in this assessment, in addition to the \textit{dimensions} of participation, we also defined \textit{modes} of participation for each dimension, in a spectrum ranging from consulting, involving, collaborating, and owning to help identify intermediate levels of stakeholder agency between the extreme poles of transactional consulting and transformative empowerment (see section~\ref{modes}). 

We offer this framework as a way to aid in designing for participatory approaches in AI, especially in light of the many practical constraints participation will necessarily need to be designed around. Specifically, for researchers and practitioners aiming to move beyond a consultative mode of participation, this will require setting ambitious, yet reasonable goals based on stakeholder access, project timelines, and resources (including funding). It will also require researchers and practitioners to move away from a role in which they oversee participation, to one where they collaboratively design and conduct the participatory approach collaboratively with stakeholders. 



However, it is critical to acknowledge that some participation may \textit{not} be better than no participation, in situations where participation is extractive (and thus harms participants \cite{sloane2020participation,dourish2020iterated}), tokenistic in ways that reify or amplify existing power structures within participating communities or between stakeholders and designers \cite{cooke2001participation}, or cases where ``pseudo-participation'' \cite{palacin2020design} may foreclose or co-opt more meaningful forms of participation \cite{kelty2016toomuch,cooke2001participation}.
Much like in participatory design more broadly, truly participatory approaches to AI will take more than``\textit{add[ing] diverse users and stir[ring]}'' to achieve substantive forms of participation and it will require ``\textit{new ways of thinking and new methods of openness}'' \cite{PD-Muller-2002}.

\looseness=-1

\subsection{Unique challenges for participation in AI}

The sociotechnical complexity of AI systems raises challenges to meaningful stakeholder participation not addressed by existing frameworks of participation in other domains.
Below, we highlight specific characteristics of AI design and development that require new ways to conceptualize and operationalize stakeholder participation in AI:

\subsubsection{Between High-Dimensionality and Human Reasoning}

With the prevalence of machine learning techniques, AI systems that are operationally useful often suffer from the ``curse of dimensionality''~\cite{domingos2012few}.
Such systems speaks the language of ``mathematical optimization in high-dimensionality'' that often goes beyond ``what can be easily grasped by a reasoning human''~\cite{burrell2016machine}. 
This dilemma of interpretability is further compounded when AI systems are iteratively retrained on new data~\cite{ananny2018seeing}.
To this end, we resonate with~\citet{ananny2018seeing} that ``if a system is so complex that even those with total views into it are unable to describe its failures and successes, then accountability models might focus on whether the system is sufficiently understood---or understandable---to allow its deployment in different environments...or if the system should be built at all.''

This calls attention to how such interpretability marks the distinction between meaningful participation and mere inclusion. As \citet{Birhane2022power} put, ``Being included might have practical consequences on the ability of people and groups to engage in participatory processes. But inclusion is not necessarily participation, since any individual can be included in a group, yet not participate, e.g., by never voting, writing, or acting.''

As a starting point for meaningful participation, we call for practitioners to direct their attention toward explicating (a) the mechanisms of particular AI systems and (b) the potentials of AI broadly construed. 
This can be operationalized on three levels.
When designing AI systems, practitioners should consider ways to explain how a given AI system work as part of humans' interactions with such a system.
Much of the effort branded as \textit{Explainable AI} falls under this pursuit~\cite[e.g.,][]{liao2020questioning, kaur2022sensible, ehsan2020human}. 
During activities for stakeholder participation, practitioners should foster localized understandings of what a particular AI system in question can do and how such a system works.
And more fundamentally, practitioners should create opportunities to foster broad understandings of \textit{what AI can do} and \textit{how AI works} prior to participation.

\subsubsection{Between Scaling and Caring}

Careful human interventions and scaling up are often conceived of as naturally opposed to each other~\cite{seaver2021care, lempert2016scale, tsing2012nonscalabilitythe}.
Two decades ago, ~\citet{neumann1996making} questioned to what extent stakeholder participation could be realized in large-scale design projects. 
The scale at which AI systems operate warrants special considerations of what constitutes meaningful participation.

In the context of contemporary AI systems, a defining character of \textit{scaling} is the shift away from rule-based AI systems toward more data-driven paradigms such as deep learning~\cite{lecun2015deep}.
The nature of data collection and curation processes involved in training ML and AI models means that large numbers of stakeholders may be nominally `participating' in AI development as the `human infrastructure' that enables ML systems to work \cite{mateescu2019ai}. 
Such nominal participation may involve ``potentially indefinitely many actors'' if the system employs third-party architectures or toolkits \cite{cooper2022accountability}. 

These shifts raise questions about whether meaningful human participation is even feasible at the scales at which many AI systems operate~\cite{seaver2021care, lempert2016scale, tsing2012nonscalabilitythe, bender2021dangers, gillespie2020content, hanna2020against}---including both the spatial scales of collection of large-scale datasets and deployment of AI systems across geographic contexts~\cite{madaio2022assessing}, as well as the rapid temporal scales at which data is collected, models are trained, and systems are deployed~\cite{chilana2015user}.
As \citet{miceli2020between} and \citet{sloane2020participation} have argued, such forms of participation may be tokenistic, extractive, or may inadvertently reinforce or amplify existing power dynamics rather than challenging power structures, as in the aims of participatory design \cite{simonsen2012routledge}. 

Given the scale of many AI systems, although some teams adopt methods from HCI to obtain narrowly scoped stakeholder participation on an individual element of a larger AI assemblage (e.g., model performance metrics \cite{shen2020designing}), it is not clear how these methods map to participation in larger system-wide decisions related to large-scale data collection and storage 
or decisions about the structure and purpose of large-scale pre-trained models, such as Large Language Models (LLMs) often fine-tuned for downstream applications \cite{bommasani2021opportunities}. 
Further, designing for participation in situations where the system is constantly adapting to new training inputs raises critical questions about when and how often and by which means it makes sense to bring stakeholders into the loop \cite[e.g.,][]{lee2019webuildai,does-AI-make-PD-obsolete}.

\subsection{Risks in using proxies for participation} 

In our analysis, we observed the use of stakeholder proxies as a key tactic in a variety of participatory AI projects. A proxy-based approach to participation was often justified due to the limited time that any given stakeholder may have to fully participate as well as the limited time and resources AI teams might have to properly recruit and engage with stakeholders.
Yet, it is worth grappling seriously with the implications that the use of proxies has as a ``labor-saving device'' \cite{mulvin2021proxies} or as a form of `discount' participation (following \citet{young2019toward,borning2004designing}). 
In particular, we call attention to proxy-related challenges on three fronts---representativeness, power imbalances, and agency.


\subsubsection{Mistaken identity}

As discussed in section~\ref{proxies_participation}, stand-ins are often recruited to represent the preferences and values of a particular group that they may belong to demographically or with which they share some aspect of lived and/or work experience.  
Yet, these approaches often rely on the assumption that stakeholders acting as a proxy may be able to adequately speak on behalf of the preferences for other stakeholders in a similar role (e.g., other teachers or other museum guides), or for stakeholders in roles they work closely with (e.g., students), or for other members of the same sociolinguistic or demographic group. 
Further, often due to selection bias, these stand-in selections may run counter to the goal of including more marginalized members of communities in the AI design process \cite[cf.][]{young2019toward,abelson2003deliberations, chasalow2021representativeness}.
There is good reason to be suspicious of the validity of these assumptions, in part due to the fact that communities are not monolithic and each person's experiences, needs, and preferences may vary. 
This becomes even more fraught when stand-ins are used to provide input into value-based design or evaluations of fairness and ethics of AI systems \cite[e.g.,][]{madaio2022assessing}.

Representation on behalf of citizens has a long history in participatory democracy, including in representative deliberation, citizen assemblies, and community juries \cite{abelson2003deliberations,pena2020innovative}, approaches that are gaining increasing traction in AI and technology design, such as diverse voices panels \cite{young2019toward} and digital juries for adjudicating content moderation \cite{fan2020digital} or annotation \cite{gordon2022jury}. However, critical work on unpacking concepts such as representativeness \cite{chasalow2021representativeness} and proxies in multiple fields \cite{mulvin2021proxies} suggests that more care should be taken in examining the power dynamics involved in determining what or who constitutes a proxy \cite{mulvin2021proxies} and what is meant by such representation and representativeness, as those terms move from one discipline (e.g., politics) to another (e.g., statistics or AI design) \cite{chasalow2021representativeness}.

\subsubsection{Biased intermediation}

The practice of relying on UX/HCI professionals as mediators has long been critiqued in the PD \cite{blomberg2012ethnography,crabtree1998ethnography,mambrey1998user} and HCI \cite{bennett2019promise,costanza2020design} literature. In particular, leveraging `proxy users' or `user surrogates,' may undermine the rich experiential knowledge that participants have about their contexts and practices. 
Further, there is a risk that UX/HCI experts as stakeholder intermediaries may inadvertently elevate or augment some voices, while leaving other voices out of the conversation---as has been written about extensively in participatory democracy \cite{polletta2012freedom} and participatory development \cite{cooke2001participation}. 

\subsubsection{Stale modeling}

In the case of algorithmic models as proxies, even if these models are able to faithfully capture capture people’s preferences at the moment of training, they risk becoming stale to people's changing preferences over time. These fossilized preference models may lead to a substantive distinction between human agent preferences and algorithmic voting across policy choices \cite[cf.][]{robertson2020if}, thus functioning as what some have referred to as a `technology of de-politicization' \cite{hitzig2020normative}. 

\subsection{Limitations}
Although we developed our corpus of research articles from a range of publication venues, using a variety of keywords, we may have missed out on relevant projects in venues outside of those we included here (e.g., work published outside of computing conference venues, including HCI journals and `grey literature’ such as white papers from policy institutes), and may have missed projects that used different terms from those we searched for to describe the participatory work they did. In addition, although we conducted a purposive sampling of authors of research articles in the corpus to recruit for the interviews, there may be key perspectives of researchers and practitioners from the corpus who we may have failed to reach, or who declined to participate. 

Finally, our interviews captured the perspectives, processes, and challenges expressed by researchers and practitioners working on participation in AI. Future work should incorporate perspectives from other key stakeholders, such as policymakers or others at civil society or community-based organizations, and, most crucially of all, people who have participated in designing any of the AI systems in the corpus, or stakeholders who are or may be impacted by particular AI systems more generally, to understand whether and how \textit{they} might want to participate in shaping the design of the systems that impact their lives.

\section{Conclusion} 

In this article, we intend to provide an empirically-grounded theoretical framework to inform research and practice on participatory approaches to AI design. We surveyed the current state of participatory approaches to AI through an analysis of a corpus of relevant research and through semi-structured interviews with researchers and practitioners involved in adopting participatory approaches to AI design. We find that the current state of participatory AI leans largely on a consultative interaction with stakeholders that allows them to provide input on particular modules of AI systems, but does not integrate them as active decision makers across the life cycle of AI design projects. In this context, we find that AI researchers and practitioners struggling with a tension between their participatory ambitions and the practical constraints they face on the ground. Further, we find that AI researchers and practitioners are adopting tactics to help them negotiate these practical constraints: including relying on stand-ins to represent the interests of affected stakeholders, relying on UX expertise to bridge stakeholders and the design team, and relying on algorithmic models to elicit stakeholders' preferences---tactics that may reify or amplify existing power dynamics. 

We intend for this paper and the accompanying framework to help shed light on current practices, and to move the state of practice forward in a way that allows AI researchers and practitioners to be more deliberate and strategic in how they articulate their participatory goals and implement participatory methods on their projects.

\bibliographystyle{ACM-Reference-Format}
\bibliography{PD-AI-bibliography.bib}

\appendix
\onecolumn
\section{Corpus Description}\label{appendix:corpus}
\setlength\extrarowheight{5pt}
 { 
\small
\begin{longtable}
{L{0.15\textwidth}L{0.1\textwidth}L{0.15\textwidth}L{0.2\textwidth}L{0.15\textwidth}L{0.15\textwidth}}
\toprule
 \textbf{Paper} &\textbf{Sector} &\textbf{Domain} &\textbf{Purpose} &\textbf{Methods} &\textbf{Stakeholder}   \\\midrule
\endfirsthead
\toprule
\textbf{Paper} &\textbf{Sector} &\textbf{Domain} &\textbf{Purpose} &\textbf{Methods} &\textbf{Stakeholder}   
\endhead
 \bottomrule
\endlastfoot
Frauenberger et al. (2012) \cite{frauenberger_interpreting_2012} &Academia &Education &Scaffold the development of social skills for children &Design sessions and workshops &Children with and without Autism Spectrum Conditions \\
\hline
Duarte et al. (2014) \cite{duarte2014welcoming} &Academia &Assistive Living &Assist therapists during interventions through gesture recognition &Storytelling &Children with Autism Spectrum Disorder \\
\hline
Šabanović et al. (2015) \cite{vsabanovic2015robot} &Academia &Assistive Living &Provide companionship and therapeutic value &Design sessions and workshops &Independently living older adults experiencing co-occurring chronic mental and physical illness \\
\hline
Katan et al. (2015) \cite{katan2015interact} &Academia &Assistive Living &Foster musical creativity &Design sessions and workshops &Musicians with learning and physical disabilities \\
\hline
Vines et al. (2015) \cite{vines2015authen} &Academia &Assistive Living &Sustain peer-to-peer knowledge sharing platform &Design sessions and workshops &Older people \\
\hline
Azenkot et al. (2016) \cite{azenkot2016enabling} &Academia &Assistive Living &Interact with and guide a blind person through a building in an effective and socially acceptable way &Design sessions and workshops &People with visual impairment \\
\hline
Krishnaswamy (2017) \cite{krishnaswamy2017participatory} &Academia &Assistive Living &Increase independence of targeted stakeholder &Surveys, Prototype Evaluation &People with physical disabilities \\
\hline
Krishnaswamy, Moorthy, and Oates (2017) \cite{krishnaswamy2017preliminary} &Academia &Assistive Living &Increase independence of targeted stakeholder &Surveys, Prototype Evaluation &People with reduced motor functionality \\
\hline
Lee et al. (2017) \cite{lee2017steps} &Academia &Assistive Living &Establish social relations with people via robots &Interviews, Design sessions and workshops &Older adults, Clinical and caregiving staff \\
\hline
Rose and Björling (2017) \cite{rose2017designing} &Academia &Assistive Living  &Measure stress &Design sessions and workshops, Prototype Evaluation &Teens \\
\hline
Noothigattu et al. (2018) \cite{noothigattu2018voting} &Academia &Autonomous Vehicle &Automate ethical decisions &Crowdsource preferences &None specified \\
\hline
Luhtala et al. (2018) \cite{luhtala2018proactive} &Government &Public Services &Predictive rescue work service &Ethnography, Design sessions and workshops, Wizard of Oz, Prototype Evaluation &Personnel of rescue departments (firefighters, operators, etc.) \\
\hline
Woodruff et al. (2018) \cite{woodruff2018qualitative} &Industry &Not Domain-Specific &Explore how traditionally marginalized populations feel about algorithmic fairness &Design sessions and workshops, Interviews &Black or African American, Hispanic or Latinx, and low socioeconomic status participants in the United States \\
\hline
Woodward et al. (2018) \cite{woodward2018using} &Academia &Assistive Living &Improve recognition and accuracy &Design sessions and workshops &Older adults with depression, Therapists, Case workers \\
\hline
Chandrasekharan et al. (2019) \cite{chandrasekharan2019crossmod} &Academia &Online Moderation &Automate decisions for content moderation &Prototype Evaluation, Interviews &Moderators of online communities \\
\hline
Herbig et al. (2019) \cite{herbigmulti2019} &Academia &Translation &Translate text &Elicitation Study, Interviews &Professional Translators \\
\hline
Lee et al. (2019) \cite{lee2019webuildai} &Academia &Social Welfare &Allocate resources &Preference Aggregation &Donors, Volunteers, Recipient organizations, and Nonprofit organizations \\
\hline
Metatla, Oldfield, and Ahmed (2019) \cite{metatla2019voice} &Academia &Education &Support inclusive education &Focus Groups, Bodystorming, Design sessions and workshops &Pupils with visual impairments \\
\hline
Hitron et al. (2019) \cite{hitron2019can} &Academia &Education &Recognize gestures to support learning &Prototype Evaluation, Controlled Experiment &Children \\
\hline
Kayacik et al. (2019) \cite{kayacik2019identifying} &Industry &Arts \& Entertainment &Foster creativity for musicians &Interviews, Design sessions and workshops, Prototype Evaluation &Musicians \\
\hline
Bhargava et al. (2019) \cite{bhargava2019gobo} &Academia &Online Moderation &Filtering Content &Survey, Prototype Evaluation &Social media users \\
\hline
Grönvall and Lundberg (2019) \cite{gronvall2019pycipedia} &Academia &Public Services &Support parents with intellectual disabilities taking care of their small children &Focus Groups, Design sessions and workshops &Social workers \\
\hline
Walsh and Wronsky (2019) \cite{walsh2019novel} &Academia &Not Domain-Specific &Create more inclusive co-design experiences for marginalized populations &Wizard of Oz, Prototype Evaluation &Marginalized populations \\
\hline
Brown et al. (2019) \cite{brown2019accountability} &Government &Public Services &Determine whether children should be placed in the child welfare system &Design sessions and workshops &Families involved in the child welfare system, Employees of child welfare agencies \\
\hline
Yadav et al. (2019) \cite{yadav2019breastfeeding} &Academia &Education &Educate women in India on pregnancy-related issues &Wizard of Oz &Breastfeeding mothers from the slum areas of Delhi, India \\
\hline
Møller, Shklovski, and Hildebrandt (2020) \cite{holten2020shifting} &Academia &Public Services &Support decision-making for job placement &Design sessions and workshops &Case workers \\
\hline
Alfrink and Tur (2020) \cite{alfrink2020contestable} &Academia &Urban Planning &Make decisions for electric vehicle (EV) charge points &Prototype Evaluation &Electric vehicle drivers \\
\hline
Candello et al. (2020) \cite{candello2020co} &Industry &Education &Teach children basic AI concepts &Role-Playing, Design sessions and workshops &Museum curators, guides, and educators \\
\hline
Kearns et al. (2020) \cite{kearns2020wizard} &Academia &Healthcare &Communicate health information &Wizard of Oz, Design sessions and workshops &Clinicians \\
\hline
Scott et al. (2022) \cite{scott2022algorithmic} &Academia &Public Services &Support job counselors in assessing job seekers as well as resource allocation (e.g., skills training, unemployment benefits) &Design Sessions/Workshops & Job seekers, practitioners \\
\hline
Mucha et al. (2020) \cite{mucha2020co} &Academia &Multiple Domains &Involve non-experts in complex technical discourses &Workbook Sprint &University students as non-experts \\
\hline
Saxena and Guha (2020) \cite{saxena2020participatory} &Academia &Public Services &Predict if a child needs state assistance &Ethnography, Interviews &Case workers in the Child-Welfare System \\
\hline
Skinner, Brown, and Walsh (2020) \cite{skinner2020children} &Academia &Education &Function as librarian in local libraries &Design sessions and workshops &Children of color in non-affluent neighborhoods of Baltimore City \\
\hline
Smith et al. (2020) \cite{smith2020keeping} &Academia, Nonprofit &Online Moderation &Predict quality of edits &Interviews &Wikipedia editors \\
\hline
Troiano, Wood, and Harteveld et al. (2020) \cite{troiano_and_2020} &Academia &Intimacy &Foster sexual interactions with humans &Story Completion Method (SCM) &Communities of fluent English writers (e.g., "Fantasy Writers" on Reddit) \\
\hline
Yu et al. (2020) \cite{yu2020keeping} &Academia &Criminal Justice &Criminal defendants' likelihood to re-offend &Prototype Evaluation, Experimental Study, Interviews &Criminal justice experts \\
\hline
Georgiou et al. (2020) \cite{georgiou2020applying} &Academia &Assistive Living &Support stroke survivors in the home environment &Sketching, Demoing, Focus Groups &Stroke survivors \\
\hline
Halfaker and Geiger (2020) \cite{halfaker2020ores} &Academia, Nonprofit &Online Moderation &Predict quality of edits &Aggregate stakeholder inputs on the design of system specs &Wikipedia editors \\
\hline
Zhang, Bellamy, and Varshney (2020) \cite{zhang2020joint} &Industry &Not Domain-Specific &Support decision-making &Preference Aggregation &None specified \\
\hline
Sendak et al. (2020) \cite{sendak2020real} &Academia &Healthcare &Diagnose conditions &Immerse project team in existing workflow, Accept project proposal from stakeholders &Front-line clinical professionals \\
\hline
Cheng et al. (2020) \cite{cheng2021soliciting} &Academia &Public Services &Predict child maltreatment &Preference Aggregation &Social workers, Parents \\
\hline
Park and Lim (2020) \cite{park_investigating_2020} &Academia &Communication &Foster family-oriented sociality &Design sessions and workshops &People in family units \\
\hline
Vaziri et al. (2020) \cite{vaziri_exploring_2020} &Academia &Service Operation &Collaborate with humans for service work &Design sessions and workshops &Service operators \\
\hline
Katell et al. (2020) \cite{katell2020facct} &Academia &Public Policy &Re-center power with those most disparately affected by the harms of algorithmic systems &Design sessions and workshops &Non-profit organizations \\
\hline
Lin and Van Brummelen (2021) \cite{lin_engaging_2021} &Academia &Education &Integrate AI into core curriculum to leverage learners’ interests &Design sessions and workshops &K-12 teachers \\
\hline
Neto, Nicolau, and Paiva (2021) \cite{neto_community_2021} &Academia &Education / Assistive Living &Foster collaborative learning for visually impaired children &Contextual inquiries, Participant observation, Interviews &Visually impaired children \\
\hline
Jacobs et al. (2021) \cite{jacobs_designing_2021} &Academia &Healthcare &Support antidepressant treatment decisionsa &Interviews, Focus groups, Prototype evaluation &Primary care physicians \\
\hline
Heitlinger et al. (2021) \cite{heitlinger_algorithmic_2021} &Academia &Agriculture &Foster decentralised non-extractive value exchange &Role playing, Design sessions and workshops &Urban agricultural communities \\
\hline
Subramonyam, Seifert, and Adar (2021) \cite{subramonyam_towards_2021} &Academia &Not Domain-Specific &Identify desirable AI characteristics &Data probes &AI designers and engineers \\
\hline
Scurto, Caramiaux, and Bevilacqua (2021) \cite{scurto_prototyping_2021} &Academia &Arts \& Entertainment &Foster music performance and robotic art &Art practices &Musicians \\
\hline
Raz et al. (2021) \cite{raz2021face} &Academia &Not Domain-Specific &Demonstrate disparate accuracy rates in facial recognition &Interactive demoing, Panel discussions, &Community organizers \\
\hline
Lee et al. (2021) \cite{lee_participatory_2021} &Academia &Labor Relations &Manage the workforce &Preference aggregation &Shift workers \\
\hline
Schelenz et al. (2021) \cite{schelenz_theory_2021} &Academia &Online Moderation &Foster diversity-aware social media platform &Preference aggregation &None specified \\
\hline
Bozic Yams and Aranda Muñoz (2021) \cite{bozic2021poetics} &Academia &Labor Relations &Augment employee creativity and support their well-being at work &Art practices, Design sessions and workshops &Professionals with non-technical background \\
\hline
Li et al. (2021) \cite{li2021nbsearch} &Academia &Software Engineering &Enable natural language search queries and intuitive visualizations for code search &Brainstorming sessions, Interviews, Prototype Evaluation, Design sessions and workshops &Experts in software engineering \\
\hline
Flügge, Hildebrandt, and Møller (2021) \cite{ammitzboll_flugge_street-level_2021} &Government &Employment &Support decision-making for job placement &Design sessions and workshops, Interviews &Case workers \\
\hline
Candello, Pichiliani, and Pinhanez (2021) \cite{candello2021museum} &Industry &Education &Teach children basic AI concepts &Role-Playing, Design sessions and workshops &Museum curators, guides, and educators \\
\hline
Saxena et al. (2021) \cite{saxena2021framework} &Academia &Public Services &Predict if a child needs state assistance &Ethnography, Interviews &Case workers in the Child-Welfare System \\
\hline
Zhang et al. (2022) \cite{zhang2022storybuddy} &Academia, Industry &Education &Create interactive storytelling experiences &Interviews, Design Sessions/Workshops &Parents, children \\
\hline
Bäuerle et al. (2022) \cite{bauerle2022symphony} &Industry &Not Domain-Specific &Not System-Specific	&Interviews, Design Sessions/Workshops, Prototype Evaluation &Machine learning practitioners \\
\hline
Zhang et al. (2022) \cite{zhang2022algorithmic} &Academia &Labor Relations &Manage, organize, coordinate, and evaluate workers &Design Sessions/Workshops, Focus Groups &Rideshare drivers \\
\hline
Park et al. (2022) \cite{park2022designing} &Academia &Labor Relations &Evaluate employees’ work performance evaluation &Design Sessions/Workshops &Employees, employers/HR teams, AI/business experts \\
\hline
Stapleton et al. (2022)
\cite{stapleton2022imagining} &Academia &Public Services &Inform workers’ decision-making for child welfare &Design Sessions/Workshops &Those who have been impacted by the child welfare system, those who work in the system \\
\hline
Suresh et al. (2022) \cite{suresh2022towards} &Academia, Nonprofit &Activism &Retrieve and filter media coverage of feminicide &Prototype Evaluation, Design Sessions/Workshops, Focus Groups, Surveys	&Activists \\
\hline
Miceli et al. (2022) \cite{miceli_documenting_2022} &Academia, Nonprofit &Not Domain-Specific &Not System-Specific &Interviews, Design Sessions/Workshops &Data Workers \\	
\hline
Storms et al. (2022) \cite{storms_transparency_2022} &Academia &Content Moderation &Provide personalized content recommendation &Prototype Evaluation, Design Sessions/Workshops &News Audiences \\
\hline
Wolf and Blomberg (2019) \cite{wolf_evaluating_2019} &Industry &Computer-Assisted Cooperative Work &Support solution design &Ethnography &IT Architects \\
\hline
Park et al. (2022) \cite{park_power_2022} &Academia &Community Engagement &Select volunteers for academic conferences &Interviews &Student Volunteers (SV) \\	
\hline
Chen et al. (2022)\cite{chen_practitioners_2022} &Academia &Content Moderation &Provide personalized content recommendation	&Interviews	&Users and Practitioners \\
\hline
Lee et al. (2020) \cite{lee_co-design_2020} &Academia &Healthcare &Identify salient features for stroke rehabilitation assessment &Interviews, Prototype Evaluation &Therapists \\
\hline
Benjamin et al. (2022) \cite{benjamin_explanation_2022} &Academia &Not Domain-Specific &Not System-Specific &Prototype Evaluation, Design Sessions/Workshops &Non-ML Experts \\
\hline
Iqbal et al. (2021) \cite{iqbal_search_2021} &Academia &Computer-Assisted Cooperative Work &Retrieve information from email records &Interviews, Test Collection &Potential Donors of Personal Records \\
\hline
Gu et al. (2021) \cite{gu_lessons_2021} &Academia &Healthcare &Diagnose disease &Prototype Evaluation &Pathologists \\
\hline
Calacci and Pentland (2019) \cite{calacci_bargaining_2022} &Academia &Labor Relations &Audit changes to platform workers' pay &Prototype Evaluation &Preference Aggregation, Platform Workers
\\
\hline
Puussaar et al. (2022) \cite{puussaar_sensemystreet_2022} &Academia &Public Services &Data-driven decision-making in local and national government &Community Meeting, Design Workshops/Sessions, Ethnography & Local Residents
\\
\hline
Kaur et al. (2019) \cite{kaur_work_2022} &Academia &Employment	&Sort and screen job applications &Speculative Design &HR recruiters and job applicants
\\
\hline
Nakao and Sugano (2020) \cite{nakao_use_2020} &Academia &Accessibility &Sound recognition &Interviews, Design Workshops/Sessions & Non-Expert Deaf, Hard-of-Hearing (DHH) People
\\
\hline
Komatsu et al. (2020) \cite{komatsu_ai_2020} &Academia &Computer-Assisted Cooperative Work &Not System-Specific &Interviews &Journalists
 \\
\hline
Alvarado et al.(2020) \cite{alvarado2020foregrounding} &Academia &Not Domain-Specific &Not System-Specific &Interviews, Diary Studies, Design Workshops/Sessions &Not-specific
\\
\hline
Seidelin et al. (2020) \cite{seidelin2020co} &Academia &Not Domain-Specific &Not System-Specific &Interviews, Design Workshops/Sessions, Prototype Evaluation &Domain Experts
\\
\end{longtable}
}


\end{document}